\documentclass[runningheads]{llncs}
\usepackage[T1]{fontenc}
\usepackage{graphicx}
\usepackage{amsmath}
\usepackage{amssymb}
\usepackage{algorithm2e}
\usepackage{multirow}
\usepackage{xcolor}
\usepackage{textcomp}
\usepackage{array}
\usepackage[hidelinks]{hyperref}
\usepackage{cleveref}

\newcommand{\sharonnew}[1]{\textcolor{violet}{SS: {\em #1}}}
\newcommand{\odednew}[1]{\textcolor{blue}{OP: {\em #1}}}
\newcommand{\sharon}[1]{\textcolor{violet}{#1}}
\newcommand{\jrk}[1]{\textcolor{olive}{#1}}
\newcommand{\oded}[1]{\textcolor{blue}{#1}}
\newcommand{\alex}[1]{\textcolor{green}{#1}}

\renewcommand{\sharonnew}[1]{}
\renewcommand{\odednew}[1]{}
\renewcommand{\sharon}[1]{}
\renewcommand{\jrk}[1]{}
\renewcommand{\oded}[1]{}
\renewcommand{\alex}[1]{}

\makeatletter
\newcommand{\removelatexerror}{\let\@latex@error\@gobble}
\makeatother

\newcommand\EqLabel[1]{\refstepcounter{equation}(\theequation)\label{#1}}
\SetStartEndCondition{ }{}{}%
\SetKwProg{Fn}{def}{\string:}{}\SetKwFunction{Range}{range}%%
\SetKw{KwTo}{in}\SetKwFor{For}{for}{\string:}{}%
\SetKwIF{If}{ElseIf}{Else}{if}{:}{elif}{else:}{}%
\SetKwFor{While}{while}{:}{}%
\AlgoDontDisplayBlockMarkers\SetAlgoNoEnd\SetAlgoNoLine%

\begin{document}
%
% \title{Contribution Title\thanks{Supported by organization x.}}
%\title{Better Inductive Generalization with Separation in PDR/IC3}
%\title{Taming the Search Space Explosion for Invariants with Quantifier Alternations}
\title{Inferring Invariants with Quantifier Alternations: Taming the Search Space Explosion}
%Taming Quantifier Alternations in Invariant Inference
%
\titlerunning{Inferring Invariants with Quantifier Alternations}
\author{Jason R. Koenig\inst{1}\and Oded Padon\inst{2} \and Sharon Shoham\inst{3} \and Alex Aiken\inst{1}}
\institute{
Stanford University, Stanford CA, USA
\and
VMware Research, Palo Alto CA, USA
\and
Tel Aviv University, Tel Aviv, Israel
}

% \author{
% \IEEEauthorblockN{Jason R. Koenig}
% \IEEEauthorblockA{\textit{Stanford University} \\
% Stanford, USA \\
% jrkoenig@stanford.edu}
% \and
% \IEEEauthorblockN{Oded Padon}
% \IEEEauthorblockA{\textit{VMware Research} \\
% Palo Alto, USA \\
% oded.padon@gmail.com}
% \and
% \IEEEauthorblockN{Sharon Shoham}
% \IEEEauthorblockA{\textit{Tel Aviv University} \\
% Tel Aviv, Israel \\
% sharon.shoham@gmail.com}
% \and
% \IEEEauthorblockN{Alex Aiken}
% \IEEEauthorblockA{\textit{Stanford University} \\
% Stanford, USA \\
% aiken@cs.stanford.edu}
% }

\maketitle
\begin{abstract}
We present a PDR/IC3 algorithm for finding inductive invariants with quantifier alternations.
We tackle scalability issues that arise due to the large search space of quantified invariants by combining a breadth-first search strategy and a new syntactic form for quantifier-free bodies.
The breadth-first strategy prevents inductive generalization from getting stuck in regions of the search space that are expensive to search
and focuses instead on lemmas that are easy to discover.
%The new syntactic form is well-suited to find lemmas with quantifier alternations by allowing both limited conjunction and disjunction in the quantifier-free body, while carefully controlling the size of the search space.
The new syntactic form is well-suited to lemmas with quantifier alternations by allowing both limited conjunction and disjunction in the quantifier-free body, while carefully controlling the size of the search space.
%The breadth-first strategy combined with the new syntactic form enables us to introduce useful inductive bias by prioritizing lemmas with simpler quantifier structures and quantifier-free bodies according to both well-defined syntactic metrics and the empirically useful metric of preferring lemmas that are fast to discover.
%Combining the breadth-first strategy with the new syntactic form results in useful inductive bias by prioritizing lemmas with simple quantifier structures and quantifier-free bodies according to both well-defined syntactic metrics and the empirically useful metric of preferring lemmas that are fast to discover.
Combining the breadth-first strategy with the new syntactic form results in useful inductive bias by prioritizing lemmas according to: (i) well-defined syntactic metrics for simple quantifier structures and quantifier-free bodies, and (ii) the empirically useful heuristic of preferring lemmas that are fast to discover.
% Our evaluation focuses on quantified invariants for distributed protocols.
%On a benchmark of examples of primarily distributed protocols, we demonstrate that our algorithm can solve more of the most complicated examples, such as Paxos variants, than existing systems.
On a benchmark suite of primarily distributed protocols and complex Paxos variants, we demonstrate that our algorithm can solve more of the most complicated examples than state-of-the-art techniques.
\end{abstract}

% \begin{IEEEkeywords}
% invariant inference, quantifier alternation, PDR/IC3, distributed protocols
% \end{IEEEkeywords}

\section{Introduction}

Invariant inference is a long-standing problem in formal methods, due to the desire for verified systems without the cost of manually writing invariants. For complex unbounded systems the required invariants often involve quantifiers, including quantifier alternations. For example, an invariant for a distributed system may need to quantify over an unbounded number of nodes, messages, etc. Furthermore, it may need to nest quantifiers in alternation (between $\forall$ and $\exists$) to capture the system's correctness arguments. For example, one crucial invariant of the Paxos consensus protocol is ``every decision must come from a quorum of votes'', i.e. $\forall \text{\em decision}. \exists \text{\em quorum}. \forall \text{\em node}.\, \text{\em node} \in \text{\em quorum} \Rightarrow \text{\em node voted for decision}$. We show that automatically inferring such invariants is possible for systems beyond the current state of the art by addressing several scalability issues that arise as the complexity of systems and their invariants increases.

%\oded{I'm not sure if we should even say this, and if so where: It is possible to manually transform a transition system to eliminate existential quantifiers from the invariant, but doing so requires knowledge of the existential invariant and is error-prone. Rather, we would like a system which can infer the quantified invariants in their original forms, with alternations.}

% Increasing automation in verification comes in two forms: automation in new domains and increasing the complexity of problems than can be solved within a domain. In this work, we focus on the second\sharon{latter}, within the domain of invariants for\sharon{remove "invariants for" -- it's repetitive} systems that require quantified invariants with quantifier alternations (both $\forall$ and $\exists$).

Many recent successful invariant inference techniques, including ours, are based on PDR/IC3~\cite{IC3,PDR}.
PDR/IC3 is an algorithmic framework for finding inductive invariants \emph{incrementally}, rather than attempting to find the entire inductive invariant at once.
PDR/IC3 progresses by building a collection of \emph{lemmas}, organized into \emph{frames} labeled by number of steps from the initial states,
until eventually some of these lemmas form an inductive invariant.
% Originally inferring Boolean formulas, PDR/IC3 was extended to several SMT theories~\cite{DBLP:conf/sat/HoderB12,DBLP:journals/fmsd/KomuravelliGC16}, universally quantified first-order formulas~\cite{UPDR}, and most recently first-order formulas with quantifier alternations~\cite{PLDI20}.
% The general framework of PDR/IC3 is shared by all these systems:
% the candidate formulas of the invariant, or lemmas, are organized into \emph{frames},
% which represent what the algorithm knows about states reachable in a certain number of steps from the initial states.
New lemmas are generated by \emph{inductive generalization}, where a given (often backward reachable) state
is generalized to a formula that excludes it and is inductive \emph{relative} to a previous frame.
%Different techniques based on PDR/IC3 differ in how they realize inductive generalization.\sharon{they differ also in other dimensions, including how to select states to be blocked. Make this a bit more general. }
Inductive generalization therefore plays a key role in PDR/IC3 implementations.
Specifically, extending PDR/IC3 to a new domain of lemmas requires a suitable inductive generalization procedure.
%Within the same restrictions of a meta-invariant for the frames, the techniques based on PDR/IC3 differ in how they generate new lemmas.
%The original IC3 algorithm inferred clauses derived from specific states, and later work~\cite{} extended this to universally quantified formulas based on the diagram of these states.
%\oded{I added some more details (and also commented out some parts)}

Techniques for inductive generalization, and more broadly for generating formulas for inductive invariants, are varied,
including interpolation~\cite{DBLP:conf/cav/McMillan14},
quantifier elimination~\cite{DBLP:journals/fmsd/KomuravelliGC16},
model-based techniques~\cite{UPDR},
and syntax guided synthesis~\cite{DBLP:conf/cav/FedyukovichPMG19,DBLP:conf/vmcai/ZhangGM21}.
Almost all of these existing techniques target either quantifier-free or universally quantified invariants.
%While it is sometimes possible to manually transform a transition system to eliminate existential quantifiers~\cite{DBLP:journals/lmcs/FeldmanPISS19}, doing so is tedious and requires some knowledge of the fully quantified invariant.
While it is sometimes possible to manually transform a transition system to eliminate some of the need for quantifiers~\cite{DBLP:journals/lmcs/FeldmanPISS19}, doing so is difficult and requires some knowledge of the fully quantified invariant.

We present a system that can infer quantified invariants with alternations based on \emph{quantified separation}, which was introduced in~\cite{PLDI20}.
Roughly, a separation query asks whether there is a quantified formula, a \emph{separator}, that evaluates to true on a given set of models and to false on another given set of models.
%This prior work used separation to implement inductive generalization and described the first PDR/IC3 implementation that infers formulas with quantifier alternations, but this implementation did not scale to examples requiring invariants with many lemmas or many quantifiers in each lemma, as the search space for quantified separators explodes as the number of symbols in the vocabulary and number of quantifiers increases. By contrast, our system can automatically find invariants for complex examples such as variants of the Paxos consensus protocol.
While \cite{PLDI20} used separation (as a black box) to implement inductive generalization and described the first PDR/IC3 implementation that finds invariants with quantifier alternations, 
it did not scale to challenging protocols such as Paxos and its variants.
These protocols require invariants with many symbols and quantifiers,
and the search space for quantified separators explodes as the number of symbols in the vocabulary and number of quantifiers increases.
% While~\cite{PLDI20} used separation as a black box to implement inductive generalization and described the first PDR/IC3 implementation that finds invariants with quantifier alternations, 
% it did not scale to challenging protocols such as Paxos and its variants,
% which require invariants with many lemmas and many quantifiers in each lemma,
% since the search space for quantified separators explodes as the number of symbols in the vocabulary and number of quantifiers increases.
In contrast, this work presents a technique that can automatically find such complex invariants.

%There are two primary ways to improve the inductive generalization at the core of PDR/IC3: (i) speed up each individual query, and (ii) generate a better (more general) lemma from each query so fewer total queries are required. In this work, we tackle both goals via two strategies: the first  integrates inductive generalization with separation in breadth-first way, and the second defines a form, $k$-term pDNF, for the quantifier-free Boolean structure of the separators. As we will explore, the goals of speed and generality are more closely related than may first appear: biasing towards better lemmas can itself result in faster queries and even make subsequent queries easier.
% As the search space for quantified invariants grows with more quantifiers and more vocabulary symbols, two problems arise for inductive generalization as part of PDR/IC3:
When targeting complex invariants, there are two main challenges for inductive generalization:
(i) the run time of each individual query; and
(ii) overfitting, i.e., learning a lemma that eliminates the given state but does not advance the search for an inductive invariant.
% Notably, we find that the two problems are more closely related than may first appear:
% biasing the search towards lemmas that are quickly discovered actually reduces overfitting.
We tackle both problems via two strategies: the first integrates inductive generalization with separation in a breadth-first way, and the second defines a new form, $k$-term pDNF, for the quantifier-free Boolean structure of the separators.

% There are two primary ways to improve the inductive generalization at the core of PDR/IC3: (i) speed up each individual query, and (ii) generate a better (more general) lemma from each query so fewer total queries are required. In this work, we tackle both of these goals via two strategies: the first focuses on quantifiers by integrating inductive generalization with separation, and the second defines a form, $k$-term pDNF, for the quantifier-free Boolean structure of the separators. Both of these strategies have effects on both the performance of inductive generalization and the quality of lemmas produced.

Integrating quantified separation with inductive generalization enables us to effectively use a breadth-first rather than a depth-first search strategy for the quantifiers of potential separators: we search in multiple parts of the search space simultaneously rather than exhaustively exploring one region before moving to the next. Beyond enabling parallelism, and thus faster wall-clock times, this restructuring can change which solution is found by allowing easy-to-search regions to find a solution first. %Empirically, these easier-to-find formulas generalize better and can result in faster subsequent queries.
We find that these easier-to-find formulas generalize better (i.e., avoid overfitting).

%This dynamic reordering is layered on an explicit bias based on observing formulas likely to be part of final invariants.

% An advantage of this restructuring of the search problem is that it naturally facilitates parallelization,
% resulting in faster wall-clock times and increasing the practicality of exploring larger search spaces. This breadth-first strategy not only results in faster queries; it also changes which solution is found by changing the order in which the space of potential separators is explored. We exploit this flexibility to introduce an explicit bias towards formulas that are likely to be part of final invariants.

% Using $k$-term pDNF enables us to narrow the search space within each inductive generalization query.
% Universally quantified invariants can be split into individually quantified clauses by transformation into conjunctive normal form (CNF), as $\forall$ distributes over $\wedge$.
% Accordingly, most PDR/IC3 based techniques find invariants as conjunctions of possibly quantified clauses.
% Using multiple clauses per lemma ($k$-clause CNF) creates a significantly larger search space, impeding scalability~\cite{PLDI20}. Using ordinary disjunctive normal form (DNF) is also insufficient, so we introduce $k$-term pDNF, a class of Boolean formulas inspired by human-written invariants that avoids the problems with CNF and DNF. Many of the lemmas that arise in our evaluation and would require many clauses in CNF are only $2$-term pDNF. We modify separation to search for lemmas of this form, leading to a reduced search space compared to CNF or DNF.

Using $k$-term pDNF narrows the search space for lemmas with quantifier alternations.
Universally quantified invariants can be split into universally quantified clauses by transformation into conjunctive normal form (CNF).
Accordingly, most PDR/IC3 based techniques find invariants as conjunctions of possibly quantified clauses.
However, invariants with quantifier alternations may require conjunction inside quantified lemmas (e.g., consider $\forall x. \exists y. p(y) \wedge r(x,y)$).
Using multiple clauses per lemma ($k$-clause CNF) creates a significantly larger search space, impeding scalability. Using disjunctive normal form (DNF) suffers from the same problem. We introduce $k$-term pDNF, a class of Boolean formulas inspired by human-written invariants that allows both limited conjunction and disjunction while keeping the search space manageable. Many of the lemmas that arise in our evaluation and would require many clauses in CNF are only $2$-term pDNF. We modify separation to search for lemmas of this form, leading to a reduced search space compared to CNF or DNF, resulting in both faster inductive generalization and less overfitting.

We evaluate our technique on a benchmark suite that includes challenging distributed protocols. Inferring invariants with quantifier alternations has recently drawn significant attention, with recent works,~\cite{PLDI20,IC3PO-NFM}, presenting techniques based on PDR/IC3 that find invariants with quantifier alternations but do not scale to complex protocols such as Paxos. Very recently, \cite{NSDI21} and \cite{IC3PO-fmcad} presented enumeration-based and PDR/IC3-based techniques, respectively, which find the invariant for simple variants of Paxos, but do not scale to more complex variants.
%Our experiments show our separation-based approach can scale to Paxos and several variants unsolved by prior work.
Our experiments show that our separation-based approach 
significantly advances the state-of-the-art,
and scales to several Paxos variants which are unsolved by prior works.
We also present an ablation study that investigates the individual effect of key features of our technique.
% We also present and ablation study that investigates the individual effect of each feature of our technique.
\jrk{This is very awkward, especially given the list below ends the exact same way.}
\oded{I see the point, but I think it's fine}

%\paragraph*{Contributions} The contributions of this work are:
%\oded{We can save space by not using the paragraph title}
This work makes the following contributions:
\begin{enumerate}
\item An algorithm for inductive generalization in PDR/IC3 (\Cref{sec:parallelizing-ig}) based on quantified separation that explores the search space in a parallel, breadth-first way and thus focuses on lemmas that are easy to discover without requiring \emph{a priori} knowledge of the search space.
\item A syntactic form of lemmas ($k$-pDNF, \Cref{sec:k-pdnf}) that is well-suited for invariants with quantifier alternations.
\item A combined system (\Cref{sec:algorithm}) able to infer the invariants of challenging protocols with quantifier alternations, including complex Paxos variants.
%\item A comprehensive evaluation (\Cref{sec:evaluation}) on a large benchmark of invariant inference examples that includes challenging Paxos variants, comparisons with a variety of state-of-the-art tools, and an ablation study of the important features of our method.
\item A comprehensive evaluation (\Cref{sec:evaluation}) on a large benchmark suite including complex Paxos variants,
comparisons with a variety of state-of-the-art tools,
and an ablation study exploring the effects of key features of our technique.
\end{enumerate}

\section{Background}\label{sec:background}
% Explain separation, IC3, related work
We review first-order logic, quantified separation, the invariant inference problem, and PDR/IC3.
\paragraph{First-Order Logic.}
We consider formulas in many-sorted first-order logic with uninterpreted functions and equality. A \emph{signature} consists of a finite set of sorts and sorted constant, relation, and function symbols. A first-order \emph{structure} over a given signature consists of a \emph{universe} set of sorted elements along with interpretations for each symbol. A structure is finite when its universe is finite. We use the standard definitions for \emph{term}, \emph{atomic formula}, \emph{literal}, \emph{quantifier-free formula}.
% The interpretation for a constant is an element, and the interpretations for relations and functions is a mapping from well-sorted tuples of the appropriate arity to $\{\top, \bot\}$ or an element of the correct sort, respectively.
% A structure is \emph{finite} if its universe is finite. A \emph{term} is a variable, a constant symbol or a function symbol applied to terms. An \emph{atomic} formula is a relation or equality applied to terms. A \emph{literal} is an atomic formula or the negation of an atomic formula.
% A formula is \emph{quantifier-free} if it consists only of boolean connectives ($\vee$, $\wedge$, $\neg$, $\Rightarrow$) and literals.
\emph{Quantified} formulas may contain universal ($\forall$) and existential ($\exists$) quantifiers with sorted variables (e.g. $\forall x{:}s_1.\, p$). A formula is in \emph{prenex normal form} if it consists of a (possibly empty) quantification \emph{prefix} followed by a quantifier-free \emph{matrix}. Any formula can be mechanically transformed into an equivalent prenex formula.
A structure $M$ satisfies a formula $p$, written $M \models p$, if the formula is true when the symbols in $p$ are interpreted according to $M$ under the usual semantics. If such an $M$ exists, then $p$ is \emph{satisfiable} and $M$ is a \emph{model} of $p$.

% \subsection{Effectively Propositional Reasoning (EPR)}
% Satisfiability of a standard first-order logic formula is only semi-decidable. Effectively Propositional Reasoning (EPR, \cite{PaxosMadeEpr}) is a fragment of first-order logic in which satisfiability is decidable and satisfiable formulas always have at least one finite model. The essence of EPR is to limit function symbols to ensure only a finite number of terms can be formed, and thus a finite number of model elements need to be considered. The successor function $s$ of the naturals, for example, violates this as we can form $s(s(s(\ldots)))$ ad infinitum. EPR requires a particular directed graph over sorts to be acyclic, where this graph has directed edge from each domain sort to the codomain sort for every function symbol. We also need to include edges from Skolem functions of existential quantifiers, i.e. for $\forall x{:}A. ( \cdots \exists y{:} B \cdots)$ there is an edge $A\rightarrow B$.

% In addition to functions from the signature, we also need to include edges for functions that result from Skolemization of the quantifiers in the formula. In practice, this means that , we add an edge from $A$ to $B$. %Note that because transformation to prenex form changes the scope of quantification, formulas which are in EPR may no longer be after the transformation (e.g. $(\forall x{:} A.\, p(x)) \wedge (\exists y{:} A.\, q(y)) \equiv \forall x{:} A. \exists y{:} A.\, p(x) \wedge q(y)$).\sharon{why do we mention that? I thought you don't use this currently, i.e., don't try to perform the opposite transformation}

\paragraph{Quantified Separation.}
To generate candidate lemmas, we use \emph{quantified separation}~\cite{PLDI20}. Given a set of \emph{structure constraints} and a predetermined space of formulas, separation produces a separator formula $p$ from the space %\sharonnew{not sure it's clear. maybe "a predetermined space"? I would rather say that separation is parameterized by the space, but I assume this is about trying to save space.}
that satisfies the constraints, or reports UNSEP if no such $p$ exists. The constraints are either \emph{positive} (a structure $M$ where $M \models p$), \emph{negative} (a structure $M$ where $M \not\models p$) or \emph{implication} (a pair of structures $M,M'$ where $M \models p \Rightarrow M' \models p$). Separation producing prenex formulas under some assumptions (satisfied by practical examples) is NP-complete~\cite{PLDI20}, and can be solved by translation to SAT.

\paragraph{Invariant Inference.}
The invariant inference problem is to compute an \emph{inductive invariant} for a given \emph{transition system}, which shows that only \emph{safe} states are reachable from the \emph{initial} states. We consider a transition system to be a set of \emph{states} as structures over some signature satisfying an axiom $\textit{Ax}$, some initial states satisfying $\textit{Init}$, a transition formula $\textit{Tr}$ which can contain primed symbols ($x'$) representing the post-state, and safe states satisfying $\textit{Safe}$. We define \emph{bad} states as $\neg \textit{Safe}$. We define single-state implication, written $A \Rightarrow B$, as $\text{\sc Unsat}(A \wedge \textit{Ax} \wedge \neg B)$ and two-state implication across transitions, written $A \Rightarrow \text{wp}(B)$, as $\text{\sc Unsat}(A \wedge \textit{Ax} \wedge \textit{Tr} \wedge \textit{Ax}' \wedge \neg B'$).
% \begin{alignat}{2}
%     &A \Rightarrow B &&\equiv \neg(A \wedge \textit{Ax} \wedge \neg B) \label{eqn:def-implies}\\
%     &A \Rightarrow \text{wp}(B) &&\equiv \neg(A \wedge \textit{Ax} \wedge \textit{Tr} \wedge \textit{Ax}' \wedge \neg B') \label{eqn:def-relative-inductive}
% \end{alignat}
% Given a \emph{transition system}, consisting of a set of \emph{states}, \emph{transitions} between pairs of states, and subsets of \emph{initial} states and \emph{bad} states, the invariant inference problem is to produce an \emph{inductive invariant} which shows the system is \emph{safe}, i.e. that bad states are not reachable from the initial states. Concretely, we represent the state space as a signature and a set of \emph{axioms} $A_1 \ldots A_n$ that all states (structures) must satisfy; initial and bad states are represented by formulas $\textit{init}$ and $\neg \textit{safe}$, and the transitions are represented as a formula $\textit{Tr}$ over a signature augmented with primed symbols for each mutable symbol of the signature. Primed symbols represent the post-state of a transition. We can represent transitions equivalently by defining $A \Rightarrow \text{wp}(B)$ as:
% \[
%     A \Rightarrow \text{wp}(B) \equiv \neg(A \wedge \textit{Tr} \wedge \neg B')
% \]
% where $B'$ is $B$ where every mutable symbol is replaced by its primed version. This can be read as ``any transition with a pre-state satisfying $A$ must have a post-state satisfying $B$.''
An \emph{inductive invariant} is a formula $I$ satisfying:

\vspace{\abovedisplayskip}\par\noindent{\setlength\tabcolsep{0pt}
\begin{tabular}{>{\centering}p{0.29\linewidth}r>{\centering}p{0.29\linewidth}r>{\centering}p{0.3\linewidth}r}
$\textit{Init} \Rightarrow I$ & \EqLabel{eqn:inv-init} &
$I \Rightarrow \text{wp}(I)$       & \EqLabel{eqn:inv-preserve} &
$I \Rightarrow \textit{Safe}$      & \EqLabel{eqn:inv-safe}
\end{tabular}}\vspace{\belowdisplayskip}

\par\noindent Together, (\ref{eqn:inv-init}) and (\ref{eqn:inv-preserve}) mean that $I$ is
%\emph{inductive} for the system,
satisfied by all reachable states,
and (\ref{eqn:inv-safe}) ensures the system is safe. We only consider invariant inference for safe systems. 

\paragraph{PDR/IC3.}\label{sec:pdr-ic3}

% \begin{figure}[t]
% \centering \includegraphics[width=1.4in]{ic3-frames.eps}
% \caption{Illustration of lemmas ($p_i$) and frames ($F_i$) in PDR/IC3. Each lemma is recorded as being in some finite frame ({$\scriptscriptstyle\blacksquare$}) or in $F_\infty$ ($\rightarrow$). In this example, $p_2$ cannot be pushed to $F_3$ because of the transition $s\rightarrow t$, where $s \models F_2$ and $t \not\models p_2$. }
% \end{figure}

PDR/IC3 is an invariant inference algorithm first developed for finite state model checking~\cite{IC3} and later extended to various classes of infinite-state systems.
%universally quantified formulas in \cite{UPDR}.
We describe PDR/IC3 as in \cite{PttT}. PDR/IC3 maintains \emph{frames} $F_i$ as conjunctions of formulas (\emph{lemmas}) representing overapproximations of the states reachable in at most $i$ transitions from ${\it Init}$. Finite frames ($i= 0,\ldots, n$) and the frame at infinity ($i=\infty$) satisfy:

\vspace{\abovedisplayskip}\par\noindent{\setlength\tabcolsep{0pt}
\begin{tabular}{>{\centering}p{0.29\linewidth}r>{\centering}p{0.29\linewidth}r>{\centering}p{0.3\linewidth}r}
${\rm\it Init} \Rightarrow F_0$ & \EqLabel{eq:ic3-frame-init-f0} &
$F_i \Rightarrow F_{i+1}$       & \EqLabel{eq:ic3-frame-subset} &
$F_n \Rightarrow F_\infty$      & \EqLabel{eq:f-n-implies-f-inf} \\[3pt]
$F_i \Rightarrow \text{wp}(F_{i+1})$ & \EqLabel{eq:f-i-implies-wp-f-ip1} &
$F_\infty \Rightarrow \text{wp}(F_\infty)$ & \EqLabel{eq:ic3-frame-rel-inductive-finf}
\end{tabular}}\vspace{\belowdisplayskip}

% \begin{tabular}{p{0.28\linewidth}p{0.36\linewidth}p{0.36\linewidth}}
% \begin{equation}%
% F_i \Rightarrow F_{i+1} \label{eq:ic3-frame-subset}%
% \end{equation} & %
% \begin{equation}%
%     F_\infty \Rightarrow \text{wp}(F_\infty) \label{eq:ic3-frame-rel-inductive-finf}%
% \end{equation} & %
% \end{tabular}}

% \noindent\begin{minipage}[t]{0.5\linewidth}%
% \begin{align}
% \\
% F_i &\Rightarrow F_{i+1} \label{eq:ic3-frame-subset} \\
% F_n &\Rightarrow F_\infty
% \end{align}
% \end{minipage}%
% \begin{minipage}[t]{0.5\linewidth}
%     \begin{align}
%     F_i &\Rightarrow \text{wp}(F_{i+1})\\
%     F_\infty &\Rightarrow \text{wp}(F_\infty) \label{eq:ic3-frame-rel-inductive-finf}
% \end{align}
% \end{minipage}
% \vspace{\belowdisplayshortskip}

\noindent Conditions (\ref{eq:ic3-frame-init-f0}), (\ref{eq:ic3-frame-subset}), and (\ref{eq:f-n-implies-f-inf}) mean ${\it Init} \Rightarrow F_i$ for all $i$, and we ensure this by restricting frames to subsets of the prior frame, when taken as sets of lemmas. Conditions (\ref{eq:f-i-implies-wp-f-ip1}) and (\ref{eq:ic3-frame-rel-inductive-finf}) say each frame is \emph{relatively inductive} to the prior frame, except $F_\infty$ which is relatively inductive to itself and thus inductive for the system. To initialize, the algorithm adds the (conjuncts of) $\it Init$ and $\it Safe$ as lemmas to $F_0$.
% As the frames are updated, preserving this meta-invariant ensures the frames represent over-approximations of reachability at a finite number of steps or for all reachable states.
The algorithm then proceeds by adding lemmas to frames using either \emph{pushing} or \emph{inductive generalization} while respecting this meta-invariant, gradually tightening the bounds on reachability until $F_\infty \Rightarrow {\it Safe}$. We can push a lemma $p \in F_i$ to $F_{i+1}$, provided $F_i \Rightarrow \text{wp}(p)$. When a formula is pushed, the stronger $F_{i+1}$ may permit us to push one or more other formulas, possibly recursively, and so we always push until a fixpoint is reached. Any mutually relatively inductive set of lemmas do not have a finite fixpoint, and we detect these sets (by checking for $F_i = F_{i+1}$) and move them to $F_\infty$.
%To initialize, we add all initial conditions and safety properties to $F_0$.\footnote{We first need to check that $\textit{Init} \Rightarrow \textit{Safe}$, or else the system is clearly not safe.}
%The algorithm refines the frames by adding formulas until the safety properties are all in $F_\infty$, at which point $F_\infty$ viewed conjunctively is an invariant of the system.

% After initialization, there are two primary ways to add formulas to a frame: \emph{pushing} and \emph{inductive generalization}. In pushing, we can push a formula $p \in F_i$ to the next frame by adding it to $F_{i+1}$, provided $F_i \Rightarrow \text{wp}(p)$. If a formula is pushed, then the new, stronger $F_{i+1}$ may permit us to push one or more formulas to $F_{i+2}$. Thus, whenever we push, we make sure to always push all formulas until a fixpoint is reached, possibly moving lemmas to $F_\infty$. If at any point any two finite frames are equal ($F_i = F_{i+1}$), we immediately push all formulas in these to $F_\infty$.

If the algorithm cannot push a lemma $p_a$ beyond frame $i$, there is a model of $\neg(F_i \Rightarrow \text{wp}(p_a))$, which is a transition $s \rightarrow t$ where $s \in F_i$ and $t \not\models p_a$. We call the pre-state $s$ a \emph{pushing preventer} of $p_a$. To generate new lemmas, we \emph{block} the pushing preventer $s$ in $F_i$ by first recursively blocking all predecessors of $s$ that are still in $F_{i-1}$, and then using an inductive generalization (IG) query to \emph{learn} a new lemma that eliminates $s$. An IG query finds a formula $p$ satisfying:

\vspace{\abovedisplayskip}\par\noindent{\setlength\tabcolsep{0pt}
\begin{tabular}{>{\centering}p{0.25\linewidth}r>{\centering}p{0.25\linewidth}r>{\centering}p{0.35\linewidth}r}
$s \not\models p$ & \EqLabel{eqn:p-negative-constraint} &
${\rm\it init} \Rightarrow p$       & \EqLabel{eqn:init-implies-p} &
$F_{i-1} \wedge p \Rightarrow \text{wp}(p)$      & \EqLabel{eqn:p-relative-inductive}
\end{tabular}}\vspace{\belowdisplayskip}

% \par\noindent{\setlength\tabcolsep{0pt}\begin{tabular}{p{0.3\linewidth}p{0.3\linewidth}p{0.4\linewidth}}
% \begin{equation}
% s \not\models p
% \end{equation} & %
% \begin{equation}
% {\rm\it init} \Rightarrow p  \label{eqn:init-implies-p}
% \end{equation} & %
% \begin{equation}
% F_{i-1} \wedge p \Rightarrow \text{wp}(p) \label{eqn:p-relative-inductive}
% \end{equation}
% \end{tabular}}\par%

% \begin{minipage}[t]{0.3\linewidth}\begin{align}\end{align}\end{minipage}%
% \begin{minipage}[t]{0.3\linewidth}\begin{align}{\rm\it init} \Rightarrow p  \label{eqn:init-implies-p}\end{align}\end{minipage}%
% \begin{minipage}[t]{0.4\linewidth}\begin{align}\end{align}\end{minipage}%

% \begin{align}
% & \\
% &{\rm\it init} \Rightarrow p  \label{eqn:init-implies-p}\\
% &F_{i-1} \wedge p \Rightarrow \text{wp}(p) \label{eqn:p-relative-inductive}
% \end{align}
If we can learn such a lemma, it can be added to $F_i$ and all previous frames, and removes at least the state $s$ stopping $p_a$ from being pushed. Classic PDR/IC3 always chooses to block the pushing preventer of a safety property (lemma from $\it Safe$) or a predecessor thereof, but other strategies have been considered \cite{PttT}.
The technique used to solve IG queries controls what kind of invariants we are able to discover. In this work we use separation to solve for $p$, which lets us infer invariants with quantifier alternations.

\section{Breadth-First Inductive Generalization with Separation}\label{sec:parallelizing-ig}
% \begin{figure}[t]
% \centering\includegraphics[width=3.45in]{refinement-loops.eps}
% \caption{The three refinement loops of our proposed algorithm. The PDR/IC3 provides states to block (along with a prior frame), which are converted into structure constraints, which become clauses in a SAT query. In the reverse direction, SAT assignments become separators, which become new lemmas in a PDR/IC3 frame. Each loop continues until a correct solution is found or the constraints are unsatisfiable.\label{fig:refinement-loops}}
% \end{figure}

% Inductive generalization is the core of PDR/IC3, and making it faster for difficult queries is a prime concern. We can view our algorithm as a set of three nested refinement loops, as depicted in \Cref{fig:refinement-loops}, where inductive generalization is the central loop. Our technique for inductive generalization turns the states to block from PDR/IC3 into a series of separation queries over a growing set of structure constraints. The separation procedure then contains a refinement loop that produces clauses for SAT queries.

Inductive generalization is the core of PDR/IC3, and improving it comes in two flavors: making individual queries faster, and generating better lemmas that are more general. We address both of these concerns by restructuring the search to be \emph{breadth-first} rather than \emph{depth-first}. We first discuss naively solving an IG query with separation (as in \cite{PLDI20}), then present an algorithm that restructures the search in a breadth-first manner.
% Here we are discussing the \emph{mechanism} by which we can divide and prioritize the search space, which allows us to discuss the \emph{policy} of what specific kinds of formulas to prioritize in \Cref{sec:k-pdnf}.

\subsection{Naive Inductive Generalization with Separation}
An IG query is solved in \cite{PLDI20} with separation by a simple refinement loop, which performs a series of separation queries with an incrementally growing set of structure constraints. Starting with a negative constraint $s$ for the state to block, we ask for a separator $p$ and check if \cref{eqn:init-implies-p,eqn:p-relative-inductive} hold for $p$ using a standard SMT solver. If both hold, $p$ is a solution to the IG query. Otherwise, the SMT solver produces a model which becomes either a positive constraint (corresponding to an initial state $p$ violates) or an implication constraint (a transition edge that shows $p$ is not relatively inductive to $F_{i-1}$), respectively.

At a high level, the SAT-based algorithm for separation from~\cite{PLDI20} uses Boolean variables to encode the kind ($\forall$/$\exists$) and sort of each quantifier, and additional variables for the presence of each syntactically valid literal in each clause in the matrix, which is in CNF. It then translates each structure constraint into a Boolean formula over these variables such that satisfying assignments encode formulas with the correct truth value for each structure. The details of the translation to SAT are not relevant here, except a few key points: (i)~separation considers each potential quantifier prefix essentially independently, (ii)~complex IG queries can result in hundreds or thousands of constraints, and (iii)~prefixes, as partitions of the space of possible separators, vary greatly in how quickly they can be explored. Further, with the black box approach where the prefixes are considered internally by the separation algorithm, even if the separation algorithm uses internal parallelism as suggested in \cite{PLDI20}, there is still a serialization step when a new constraint is required. As a consequence of (ii) and (iii), a significant failure mode of this naive approach is that the search becomes stuck generating more and more constraints for difficult parts of the search space that ultimately do not contain an easy-to-discover solution to the IG query.
% \sharonnew{even if this is really all that one needs to know to understand what we're doing, I think the readers will get the feeling that they are missing important details. Can we give a brief description of the separation algorithm? Something like: At a high level, the separation algorithm attaches Boolean variables to each unknown quantifier in the quantifier prefix of the separator, determining its identity ($\forall, \exists$) and sort. It further attaches Boolean variables to each potential literal and clause in the unknown matrix of the separator. The algorithm then encodes the structure constraints as SAT constraints over these Boolean Variables, such that a satisfying assignment induces a separator. While the details of the encoding and implementation are not relevant to our discussion, a few key details.... (I don't think it's clear that the alg must consider each quantifier prefix separately -- I think we had an example for that. Can we bring it back?)}

\subsection{Prefix Search at the Inductive Generalization Level}
\label{sec:prefixes}
To fix the problems with the naive approach, we propose lifting the choice of prefix to the IG level, partitioning a single large separation query into a query for each prefix. Each sub-query can be explored in parallel, and each can proceed independently by querying for new constraints (using \cref{eqn:init-implies-p,eqn:p-relative-inductive} as before) %\sharonnew{I think it is worth explaining that new constraints arrive just like before by checking relative inductiveness of the proposed separator. The difference is that this is done in parallel to other prefixes still being explored within the same IG query}
without serializing by waiting for other prefixes.
We call this a \emph{breadth-first} search, because the algorithm can spend approximately equal time on many parts of the search space, instead of a \emph{depth-first} search which exhausts all possibilities in one region before moving on to the next. When regions have greatly varying times to search, the breadth-first approach prevents expensive regions from blocking the search in cheaper regions.
This improvement relies on changing the division between separation and inductive generalization: without the knowledge of the formulas (\cref{eqn:init-implies-p,eqn:p-relative-inductive}) that generate constraints, the separation algorithm cannot generate new constraints on its own. %\jrk{keep this sentence?}\sharonnew{I don't like the use of "interface". I think it sounds a bit technical and low level, but I like the message and I think it is good to emphasize that this cannot be done without the integration because the separator cannot generate new constraints}

A complicating factor is that in addition to prefixes varying in difficulty, sometimes there are entire classes of prefixes that are difficult. For example, many IG queries have desirable universal-only solutions, but spend a long time searching for separators with alternations, as there are far more distinct prefixes with alternations than those with only universals.
To address this problem, we define possibly overlapping sets of prefixes, called \emph{prefix categories}, and ensure the algorithm spends approximately equal time searching for solutions in each category
(e.g., universally quantified invariants, invariants with at most one alternation and at most one repeated sort).
Within each category, we order prefixes to further bias towards likely solutions: first by smallest quantifier depth, then fewest alternations, then those that start with a universal, and finally by smallest number of existentials. 

\subsection{Algorithm for Inductive Generalization}
\begin{figure}[t]
% {\small
% \begin{verbatim}
% N # number of parallel threads
% def IG(s):
%     prefixes = (queue of prefixes respecting logic)
%     constraints = {Negative(s)}
%     for i = 0..N in parallel:
%         return worker(prefixes, constraints)
% def worker(prefixes, constraints):
%     while True:
%         pr = get_prefix_restrictions()
%         get unsolved prefix from prefixes that satisfies pr
%         sep = SeparationAlg(prefix)
%         sep.add_constraint(constraints)
%         while True:
%             candidate_p = sep.separate()
%             if candidate_p is UNSEP:
%                 mark prefix UNSEP
%                 continue
%             for constraint in related_constraints(prefix):
%                 if candidate_p does not satisfy constraint:
%                     sep.add_constraint(constraint)
%                     break
%             if added constraint: continue
%             constraint = relative_inductive_check(candidate_p)
%             if constraint is None:
%                 return candidate_p # solve IG query
%             sep.add_constraint(constraint)

% \end{verbatim}}
\removelatexerror
\begin{algorithm}[H]\scriptsize
\Fn(){\upshape IG($s$: state, $i$: frame)}{
$\forall P.\; C(P) = \{\text{Negative}(s)\}$\;
\For{\upshape $i = 1 \ldots N$ \textbf{in parallel}}{
    \While{\upshape true}{
        $P$ = next-prefix()\;
        \While{\upshape true}{
            $p$ = separate $C(P)$\;
            \uIf{\upshape $p$ is UNSEP}{break}
            \uElseIf{\upshape any $c\in R_C(P)$ and $p \not\models c$}{add $c$ to $C(P)$}
            \uElseIf{\upshape ($c :={}$SMT check \cref{eqn:init-implies-p,eqn:p-relative-inductive}) $\not=$ UNSAT}{add $c$ to $C(P)$}
            \uElse{\Return $p$ as solution}
        }
    }
}
}
\end{algorithm}
\caption{Pseudocode for our proposed inductive generalization algorithm.}
\label{fig:ig-algo-listing}
\end{figure}

We present our algorithm for IG using separation in Figure~\ref{fig:ig-algo-listing}. Our algorithm has a fixed number $N$ of worker threads which take prefixes from a queue subject to prefix restrictions, and perform a separation query with that prefix. 
Each worker thread calls next-prefix() to obtain the next prefix to consider, according to the order discussed in the previous section.
To solve a prefix $P$, a worker performs a refinement loop as in the naive algorithm, building a set of constraints $C(P)$ until a solution to the IG query is discovered or separation reports UNSEP.
%To complete the presentation, we discuss details of handling constraints efficiently.

% \paragraph{Related Constraints}
While we take steps to make SMT queries for new constraints as fast as possible (\Cref{sec:smt-robustness}), these queries are still expensive and we thus want to re-use constraints between prefixes where it is beneficial. Re-using every constraint discovered so far is not a good strategy as the cost of checking upwards of hundreds of constraints for every candidate separator is not justified by how frequently they actually constrain the search. Instead, we track a set of \emph{related constraints} for a prefix $P$, $R_C(P)$. We define related constraints in terms of \emph{immediate sub-prefixes} of $P$, written $S(P)$,
%$P' \sqsubset P$, 
which are prefixes obtained by dropping exactly one quantifier from $P$, i.e. the quantifiers of $P'\in S(P)$ are a subsequence of those in $P$ with one missing. We then define $R_C(P) = \cup_{P' \in S(P)} C(P')$, i.e. the related constraints of $P$ are all those used by immediate sub-prefixes.
While $S(P)$ considers only immediate sub-prefixes,
constraints may propagate from non-immediate sub-prefixes as the algorithm progresses.

Constraints from sub-prefixes are used because the possible separators for those queries are also possible separators for the larger prefix. Thus the set of constraints from sub-prefixes will definitely eliminate some potential separators, and in the usual case where the sub-prefixes have converged to UNSEP, will rule out an entire section of the search space. We also opportunistically make use of known constraints for the same prefix generated in prior IG queries, as long as those constraints still satisfy the current frame.

Overall, the algorithm in \Cref{fig:ig-algo-listing} uses parallelism across prefixes to generate independent separation queries in a breadth-first way, while carefully sharing only useful constraints.
From the perspective of the global search for an inductive invariant
the algorithm introduces two forms of inductive bias:
(i)~explicit bias arising from controlling the order and form of prefixes (\Cref{sec:prefixes}), and (ii)~implicit bias towards formulas which are easy to discover.

\section{\texorpdfstring{$k$}{k}-Term Pseudo-DNF}\label{sec:k-pdnf}
% So far we have considered the search strategy over quantifier prefixes.
We now consider the search space for quantifier-free matrices,
and introduce a syntactic form that shrinks the search space while still allowing common invariants with quantifier alternations to be expressed.
%Inspired by human-written invariants, we pick a form for the matrix which shrinks the search space while still allowing common invariants to be expressed.

% \subsection{\texorpdfstring{$k$}{k}-Term Pseudo-DNF}\sharonnew{strange to have just one subsection. Maybe turn it into a paragraph title instead?}
Conjunctive and disjunctive normal forms (CNF and DNF) are formulas that consist of a conjunction of \emph{clauses} (CNF) or a disjunction of \emph{cubes} (DNF), where clauses and cubes are disjunctions and conjunctions of literals, respectively: For example, $(a \vee b \vee \neg c) \wedge (b \vee c)$ is in CNF and $(a \wedge \neg c) \vee (\neg a \wedge b)$ is in DNF. We further define $k$-clause CNF and $k$-term DNF as formulas with at most $k$ clauses and cubes, respectively.

In~\cite{PLDI20} separation is performed by finding a matrix in $k$-clause CNF, biasing the search by minimizing the sum of the number of quantifiers and $k$. %Further, each clause is minimized to ensure there are no extraneous literals.
We find that both CNF and DNF are not good fits for the formulas in human-written invariants. For example,
consider the following formula from Paxos:
% \begin{figure}[t]
% \centering
%     \caption{Example formulas from our benchmark, in 2-term pDNF and 3-term pDNF, respectively.}
%     \label{fig:pdnf-examples}
% \end{figure}
% Most universal-only human-written invariants look like $\forall\forall\forall.\,  a \wedge b \rightarrow c$. When there are existentials involved, the condition is often $\forall\forall\forall.\, a \wedge b \rightarrow \exists.\,  c \wedge d \wedge e$, i.e. for every thing satisfying $a \wedge b$, there exists something satisfying $c \wedge d \wedge e$. An example of such a formula from Paxos is:
\begin{align*}\label{eqn:paxos-forall-exists-ex}
\begin{split}\forall r_1,r_2{,}v_1{,}v_2{,}q.\exists n. r_1 < r_2 \wedge \text{proposal}(r_2,v_2) \wedge v_1 \neq v_2 \\
\rightarrow \text{member}(n,q) \wedge \text{left-round}(n,r_1) \wedge \neg\text{vote}(n,r_1,v_1)
\end{split}\end{align*}
To write this in CNF, we need to distribute the antecedent over the conjunction, obtaining the 3-clause formula:
\begin{align*}
    &(r_1 < r_2 \wedge \text{proposal}(r_2,v_2) \wedge v_1 \neq v_2 \rightarrow \text{member}(n,q)) \wedge {} \\
    &(r_1 < r_2 \wedge \text{proposal}(r_2,v_2) \wedge v_1 \neq v_2 \rightarrow \text{left-round}(n,r_1)) \wedge {} \\
    & (r_1 < r_2 \wedge \text{proposal}(r_2,v_2) \wedge v_1 \neq v_2 \rightarrow \neg\text{vote}(n,r_1,v_1))
\end{align*}

% A formula with only universal quantifiers can always be factored into a conjunction of clausal formula by distribution, and each of these conjuncts can be learned incrementally by PDR/IC3. When formulas involve quantifier alternation, it is not sufficient to only consider clausal formula, as existentials do not factor across conjunction.
% Representing the matrix of (\ref{eqn:paxos-forall-exists-ex}) in CNF would require three clauses, one for each literal in the existentially quantified cube:

% sharon{1. why not also write the clauses explicitly? 2. at the risk of interrupting the flow of the example, I think this is a good place to contrast it with universal quantification, where you could split to 3 quantified clauses (i.e., restrict matrices to one clause). Can even demonstrate it on the example }

When written without $\rightarrow$, this matrix has the form $\neg a \vee \neg b \vee  c \vee (d \wedge e \wedge \neg f)$, which is already in DNF. Under the $k$-term DNF, however, the formula requires a single-literal cube for each antecedent literal, i.e. $k=4$. Because of the quantifier alternation, we cannot split this formula into cubes or clauses, and so a search over either CNF or DNF must consider a significantly larger search space.% \sharon{the bottom line is not stated: due to the quantifier alternation, we cannot split these into individual quantified clauses or cubes; this means that if we look for invariants in k-CNF or k-DNF, the effective search space must include lemmas whose matrix is taken from this significantly larger search space. (intro explained it well)}
To solve these issues, we define a variant of DNF, $k$-term pseudo-DNF ($k$-pDNF), where one cube is negated, yielding as many individual literals as needed:

\begin{definition}[$k$-term pseudo-DNF]
A quantifier-free formula $\varphi$ is in $k$-term pseudo-DNF for $k \ge 1$ if $\varphi \equiv \neg c_1 \lor c_2 \lor \ldots \lor c_k$, where $c_1,\ldots,c_k$ are cubes. Equivalently, $\varphi$ is in $k$-term pDNF if there exists $n \geq 0$ such that $\varphi \equiv \ell_1 \lor \ldots \lor \ell_n \lor c_2 \lor \ldots \lor c_k$, where $\ell_1,\ldots,\ell_n$ are literals and $c_2,\ldots,c_k$ are cubes.
\end{definition}

Note that $1$-term pDNF is equivalent to $1$-clause CNF, i.e. a single clause.
$2$-term pDNF correspond to formulas of the form $(\text{cube}) \rightarrow (\text{cube})$. Such formulas are sufficient for all but a handful of the lemmas required for invariants in our benchmark suite. An exception is the following, which has one free literal and two cubes (so it is $3$-term pDNF):
\begin{align*}
    &\forall v_1. \;\exists n_1, n_2, n_3, v_2, v_3.\; & \\
    &\quad (d(v_1) \rightarrow \neg m(n_1) \wedge u(n_1,v_1)) \vee {} &  \\
    &\quad \left(\neg m(n_2) \wedge \neg m(n_3) \wedge u(n_2,v_2) \wedge u(n_3,v_3) \wedge v_2 \neq v_3\right) &
\end{align*}

For a fixed $k$, $k$-clause CNF, $k$-term DNF, and $k$-term pDNF
all have the same-size search space, as the SAT query inside the separation algorithm will have one indicator variable for each possible literal in each clause or cube. The advantage of pDNF is that it can express more invariant lemmas with a small $k$, reducing the size of the search space while still being expressive. We can also see pDNF as a compromise between CNF and DNF, and we find that pDNF is a better fit to the matrices of invariants with quantifier alternation.

\section{An Algorithm for Invariant Inference}\label{sec:algorithm}

We now take a step back to consider the
% overall PDR/IC3 level
high-level PDR/IC3 structure
of our algorithm. We have described how our algorithm performs inductive generalization (\Cref{sec:parallelizing-ig,sec:k-pdnf}), which is the central ingredient. %\sharon{to better tie this together, I suggest to add something like "We have already described the way in which our algorithm performs inductive generalization (\Cref{sec:parallelizing-ig,sec:inductive-bias}), which is the most central ingredient." then replace "First" by "Next"}
As in all PDR/IC3 variants, we use IG to block backward reachable states (i.e., states from which a safety violation is reachable).
We next discuss blocking states that are not backward reachable as a heuristic for finding additional useful lemmas.
We then discuss how we can search for formulas in the EPR logic fragment and techniques to increase the robustness of SMT solvers. Finally, we tie everything together to give a complete description of our proposed algorithm.

\subsection{May-proof-obligations}

In classic PDR/IC3, the choice of pushing preventer to block is always that of a safety property. \cite{PttT} proposed a heuristic that in our terminology is to block the pushing preventer of other existing lemmas, under the heuristic assumption that current lemmas in lower frames are part of the final invariant but lack a supporting lemma to make them inductive. The classic blocked states are known as \emph{must-proof-obligations}, as they are states that must be eliminated somehow to prove the safety property. In contrast, these heuristic states are \emph{may-proof-obligations}, as they may or may not be necessary to block. Our algorithm selects these lemmas at random, biased towards lemmas with smaller matrices.

To block a state, we first recursively block its predecessors in the prior frame, if they exist. For may-proof-obligations,\footnote{For unsafe transition systems, this can also occur for must-proof-obligations.} this recursion can potentially reach all the way to an initial state in $F_0$, and thus proves that the entire chain of states is reachable--- i.e., the states cannot be blocked. This fact shows that the original lemma is not part of any final invariant and cannot be pushed past its current frame; it also provides a positive structure constraint useful for future IG queries.

\subsection{Multi-block Generalization}

After an IG query blocking state $s$ is successful, the resulting lemma $p$ may cause the original lemma that created $s$ to be pushed to the next frame. If not, there will be a new pushing preventer $s'$. If $s'$ is in the same frame, we can ask whether there is a single IG solution formula $p_1$ which blocks both $s$ and $s'$. %As observed in \cite{DBLP:conf/cav/KrishnanCSG20}
If we can find such a $p_1$, it is more likely to generalize past $s$ and $s'$, and  we should prefer $p_1$.
This is straightforward to do with separation: we incrementally add another negative constraint to the existing separation queries. To implement \emph{multi-block generalization}, we continue an IG query if the new pushing preventer is suitable (i.e. exists and is in the same frame), accumulating as many negative constraints as we can until we do not have a suitable state or we have spent as much time as the original query. This timeout guarantees we do not spend more than half of our time on generalization, and protects us in the case that the new set of states cannot be blocked together with a simple formula.

\subsection{Enforcing EPR}\label{sec:epr}

Effectively Propositional Reasoning (EPR, \cite{PaxosMadeEpr}) is a fragment of many-sorted first-order logic in which satisfiability is decidable and satisfiable formulas always have a finite model. The essence of EPR is to limit function symbols, both in the signature and from the Skolemization of existentials, to ensure only a finite number of ground terms can be formed. EPR ensures this property by requiring that there be no cycles in the directed graph with an edge from each domain sort to the codomain sort for every (signature and Skolem) function symbol. For example, $(\forall x{:}S.\, \varphi_1) \vee (\exists y{:}S.\, \varphi_2)$ is in EPR, but $\forall x{:}S.\, \exists y{:}S.\, \varphi_3$ is not in EPR as the Skolem function for $y$ introduces an edge from sort $S$ to itself. The acyclicity requirement means that EPR is not closed under conjunction, and so is best thought of as a property of a whole SMT query rather than of individual lemmas. 
Despite these restrictions, EPR can be used to verify complex distributed protocols~\cite{PaxosMadeEpr}.
%Some distributed protocols can be explicitly modelled such that their verification conditions are in EPR~\cite{PaxosMadeEpr}.

For invariant inference with PDR/IC3, the most straightforward way to enforce acyclicity is to decide \emph{a priori} which edges are allowed, and to not infer lemmas with disallowed Skolem edges. In practice, enforcing EPR means simply skipping prefixes during IG queries that would create disallowed edges. Without this fixed set of allowed edges, adding a lemma to a frame may prevent a necessary lemma from being added to the frame in a later iteration, as PDR/IC3 lacks a way to remove lemmas from frames. 
Requiring the set of allowed edges as input is a limitation of our technique and other state-of-the-art approaches (e.g.~\cite{NSDI21}).
% This limitation is largely orthogonal the the contributions of this paper.
%Requiring a set of allowed edges means that the user must know something of the structure of the final invariant.
We hope that future work expands the scope of decidable logic fragments, so that systems require less effort to model in such a fragment. It is also possible that our algorithm could be wrapped in an outer search over the possible acyclic sets of edges.

Because separation produces prenex formulas, some EPR formulas would be disallowed without additional effort (e.g. a prenex form of $(\forall x{:}S.\, \varphi_1) \vee (\exists y{:}S.\, \varphi_2)$ is $\forall x{:}S.\, \exists y{:}S.\, (\varphi_1) \vee (\varphi_2)$). In our implementation, we added an option where separation produces prenex formulas that may not be in EPR directly, but where the scope of the quantifiers can be \emph{pushed down} into the pDNF disjunction to obtain an EPR formula. Extra SAT variables are introduced that encode whether a particular quantified variable appears in a given disjunct, and we add the constraint that the quantifiers are nested consistently and in such a way as to be free of disallowed edges. Because this makes separation queries more difficult, we only enable this mode for the single example that requires non-prenex EPR formulas.

\subsection{SMT Robustness}\label{sec:smt-robustness}
Even with EPR restrictions, some SMT queries we generate are difficult for the SMT solvers we use (Z3~\cite{Z3} and CVC5\footnote{Successor to CVC4 \cite{CVC4}.}), sometimes taking minutes, hours, or longer. This wide variation of solving times is significant because separation, and thus IG queries, cannot make progress without a new structure constraint. We adopt several strategies to increase robustness: periodic restarts, running multiple instances of both solvers in parallel, and \emph{incremental queries}. Our incremental queries send the formulas to the SMT solver one at a time, asserting a subset of the input. An UNSAT result from a subset can be returned immediately, and a SAT result can be returned if there is no un-asserted formula violated by the model. Otherwise, one of the violating formulas is asserted, and the process repeats. This process usually avoids asserting all the discovered lemmas from a frame, which significantly speeds up many of the most difficult queries, especially those with dozens of lemmas in a frame or those not in EPR.

\subsection{Complete Algorithm}
% \begin{figure*}[t]
% \begin{minipage}[t]{0.49\textwidth}
% \removelatexerror
% \begin{algorithm}[H]\scriptsize
% \Fn(){\upshape{\sc P-Fol-Ic3}()}{
%     $F_0 = \text{init} \cup \text{safety}$\;
%     push()\;
%     \textbf{start} {\sc Learning}(), {\sc Heuristic}()\;
%     \textbf{wait} for invariant\;
% }
% \Fn(){\upshape{\sc Multiblock}($\ell$: lemma, $s$: state)}{
%     $S, i$ := \{s\}, frame of $s$\;
%     \While{\upshape not timed out}{
%         $p$ = {\sc IG}($S$, i)\;
%         speculatively add $p$ to frame $i$\;
%         $s'$ = to-block($\ell$)\;
%         \If{$s' \in F_i$}{
%             add $s'$ to $S$\;
%         }\Else{
%             add $p$ to frame $i$\;
%             push()\;
%             \Return
%         }
%     }
% }
% \end{algorithm}
% \end{minipage}\hfill%
% \begin{minipage}[t]{0.49\textwidth}
% \removelatexerror
% \begin{algorithm}[H]\scriptsize
% \Fn(){\upshape{\sc Learning}()}{
%     \While{true}{
%         $s$ = to-block(safety)\;
%         {\sc Multiblock}(safety, s)\;
%     }
% }
% \Fn(){\upshape{\sc Heuristic}()}{
%     \While{\upshape true}{
%         $\ell$ = random lemma before safety\;
%         $s$ = to-block($\ell$)\;
%         \If{$s' \in F_0$}{
%             mark $s$ reachable\;
%             mark bad lemmas\;
%         }\Else{
%             {\sc Multiblock}($\ell$, s)\;
%         }
%     }
% }
% \end{algorithm}
% \end{minipage}
% \caption{Pseudocode for our proposed inference algorithm, {\sc P-Fol-Ic3}.}\label{fig:main-algo-listing}
% \end{figure*}
\begin{figure*}[t]
\begin{minipage}[t]{0.49\textwidth}
\removelatexerror
\begin{algorithm}[H]\scriptsize
\Fn(){\upshape{\sc P-Fol-Ic3}()}{
    $F_0 = \text{init} \cup \text{safety}$\;
    push()\;
    \textbf{start} {\sc Learning}(), {\sc Heuristic}()\;
    \textbf{wait} for invariant\;
}
\Fn(){\upshape{\sc Multiblock}($\ell$: lemma, $s$: state, $i$)}{
    $S$ = \{$s$\}\;
    \While{\upshape not timed out}{
        $p$ = {\sc IG}($S$, $i$)\;
        speculatively add $p$ to frame $i$\;
        $s', i'$ = to-block($\ell$)\;
        remove $p$ from frame $i$\;
        \If{$i = i'$}{
            add $s'$ to $S$\;
        }\Else{
            \bfseries{break}
        }
    }
    add $p$ to frame $i$\;
    push()\;
}
\end{algorithm}
\end{minipage}\hfill%
\begin{minipage}[t]{0.49\textwidth}
\removelatexerror
\begin{algorithm}[H]\scriptsize
\Fn(){\upshape{\sc Learning}()}{
    \While{true}{
        $s, i$ = to-block(safety)\;
        {\sc Multiblock}(safety, $s$, $i$)\;
    }
}
\Fn(){\upshape{\sc Heuristic}()}{
    \While{\upshape true}{
        $\ell$ = random lemma before safety\;
        $s, i$ = to-block($\ell$)\;
        \If{$i = 0$}{
            mark $s$ reachable\;
            mark bad lemmas\;
        }\Else{
            {\sc Multiblock}($\ell$, $s$, $i$)\;
        }
    }
}
\end{algorithm}
\end{minipage}
\caption{Pseudocode for our proposed inference algorithm, {\sc P-Fol-Ic3}.}\label{fig:main-algo-listing}
\end{figure*}
Figure~\ref{fig:main-algo-listing} presents the pseudocode for our algorithm, which consists of two parallel tasks (learning and heuristic), each using half of the available parallelism to discharge IG queries, and pushing to fixpoint after adding any lemmas. In this listing, the $\text{to-block}(\ell)$ function computes the state and frame to perform an IG query in order to push $\ell$ (i.e. the pushing preventer of $\ell$ or a possibly multi-step predecessor thereof). The heuristic task additionally may find reachable states, and thus mark lemmas as bad. We cancel an IG query when it is solved by a lemma learned or pushed by another task.

Our algorithm is parameterized by the logic used for inductive generalization, and thus the form of the invariant. We support universal, EPR, and full first-order logic (FOL) modes. Universal mode restricts the matrices to clauses, and considers predecessors of superstructures when computing $\text{to-block}()$ (as in \cite{UPDR}). EPR mode also takes as input the set of allowed edges. In FOL mode, there are no restrictions on the prefix.

\section{Evaluation}\label{sec:evaluation}
We evaluate our algorithm and compare with prior approaches on a benchmark of invariant inference problems. We discuss the benchmark, our experimental setup, and the results.

\subsection{Invariant Inference Benchmark}
Our benchmark is composed of invariant inference problems from prior work on distributed protocols~\cite{IVy,PaxosMadeEpr,FeldmanTACAS17,PadonPOPL18,PLDI2018,CAV19-Thresholds,CAV19-mypyvy}, written in or translated to the mypyvy tool's input language \cite{Mypyvy}. Our benchmark contains a total of 30 problems (\Cref{tab:benchmark-results}), ranging from simple (toy-consensus, firewall) to complex (stoppable-paxos-epr, bosco-3t-safety). Some problems admit invariants that are purely universal, and others use universal and existential quantifiers, with some in EPR.
%The Paxos variants include both an \texttt{-forall} version and an \texttt{-epr} version, where the latter requires an invariant with quantifier alternations, and the former has been augmented with ghost state such that it can be proven with a universally quantified invariant.
All our examples are safe transition systems with a known human-written invariant.

\subsection{Experimental Setup}
We compare our algorithm to the techniques {\sc Swiss} \cite{NSDI21}, IC3PO \cite{IC3PO-NFM,IC3PO-fmcad}, fol-ic3 \cite{PLDI20}, and PDR${}^\forall$ \cite{UPDR}. We performed our experiments on a 56-thread machine with 64 GiB of RAM, with each experiment restricted to 16 hardware threads, 20GiB of RAM, and a 6 hour time limit.\footnote{Specifically, an dual-socket Intel(R) Xeon(R) CPU E5-2697 v3 @ 2.60GHz.} To account for noise caused by randomness in seed selection, we ran each algorithm 5 times and report the number of successes and the median time. PDR${}^\forall$, IC3PO, and fol-ic3 are not designed to use parallelism, while {\sc Swiss} and our technique make use of parallelism. For IC3PO, we use the better result from the two implementations \cite{IC3PO-NFM} and \cite{IC3PO-fmcad}, and give reported results for those we could not replicate. For our technique, we ran the tool in universal-only, EPR, or full FOL mode as appropriate. For $k$-pDNF, we use $k=1$ for universal prefixes and $k=3$ otherwise.
Our implementation uses five prefix categories (universal-only mode uses only the first two):
(i)~universal formulas,
(ii)~universal formulas with each sort appearing in at most two quantifiers,
(iii)~at most one quantifier alternation and each sort appearing in at most two quantifiers,
(v)~at most two quantifier alternations and each sort appearing in at most two quantifiers, and
(vi)~at most two quantifier alternations.

% PrefixConstraints(Logic.Universal),
% PrefixConstraints(Logic.Universal, max_repeated_sorts=2),
% PrefixConstraints(Logic.FOL, max_alt=1, max_repeated_sorts=2),
% PrefixConstraints(Logic.FOL, max_alt=1, max_repeated_sorts=2),
% PrefixConstraints(Logic.FOL, max_alt=2, max_repeated_sorts=2),
% PrefixConstraints(Logic.FOL, max_alt=2)]

\subsection{Results and Discussion}

\begin{table*}[t]
\caption{Experimental results, giving both the median wall-clock time (seconds) of run time and the number of trials successful, out of five. If there were less than 3 successful trials, 
we report the slowest successful trial, indicated by (>). A dash (-) indicates all trials failed or timed out after 6 hours (21600 seconds). A blank indicates no data.\label{tab:benchmark-results}}
\begin{center}
\scriptsize
%\makebox[\textwidth][c]%
{\setlength\tabcolsep{1pt}
\begin{tabular}{l|c|rrrrrrrrrrrrrrrrrr}
Example &                          {\tiny EPR} &   \bfseries{Our}      & \# & {\sc Swiss} & \# &                IC3PO & \# & fol-ic3 & \# &  PDR${}^\forall$ & \# \\\hline
lockserv &                         $\forall$   &    19                 &  5 &        9573 &  4 &                    5 &  5 &       7 &  5 &                6 &  5 \\
toy-consensus-forall &             $\forall$   &     4                 &  5 &          22 &  5 &                    4 &  5 &      11 &  5 &                4 &  5 \\
ring-id &                          $\forall$   &     7                 &  5 &         192 &  5 &                   81 &  5 &      28 &  5 &               20 &  5 \\
sharded-kv &                       $\forall$   &     8                 &  5 &       17291 &  5 &                    4 &  5 &      19 &  5 &                6 &  5 \\
ticket &                           $\forall$   &    23                 &  5 &           - &  0 &                    - &  0 &     240 &  5 &               22 &  5 \\
learning-switch &                  $\forall$   &    76                 &  5 &        1744 &  4 &                   29 &  5 &       - &  0 &               94 &  5 \\
consensus-wo-decide &              $\forall$   &    50                 &  5 &          52 &  5 &                    6 &  5 &      33 &  5 &               29 &  5 \\
consensus-forall &                 $\forall$   &  1908                 &  5 &          80 &  5 &                   15 &  5 &    1125 &  5 &              104 &  5 \\
cache &                            $\forall$   &  2492                 &  4 &           - &  0 &                 3906 &  5 &       - &  0 &             2628 &  5 \\
paxos-forall &                         $\forall$   &   885                 &  5 &           - &  0 &                    - &  0 &       - &  0 &              555 &  5 \\
flexible-paxos-forall &                $\forall$   &  1961                 &  5 &           - &  0 &                 1654 &  5 &       - &  0 &              423 &  5 \\
stoppable-paxos-forall &               $\forall$   &  7779                 &  5 &           - &  0 &                    - &  0 &       - &  0 &                - &  0 \\
fast-paxos-forall &                    $\forall$   &     -                 &  0 &           - &  0 &                    - &  0 &       - &  0 &            20176 &  3 \\
vertical-paxos-forall &                $\forall$   &     -                 &  0 &           - &  0 &                    - &  0 &       - &  0 &                - &  0 \\\hline
firewall &                         $-$         &     4                 &  5 &           - &  0 &                    3 &  5 &       9 &  5 &                  &    \\
sharded-kv-no-lost-keys &          $\checkmark$&     4                 &  5 &           9 &  5 &                    4 &  5 &       5 &  5 &                  &    \\
toy-consensus-epr &                $\checkmark$&     4                 &  5 &          10 &  5 &                    4 &  5 &      49 &  5 &                  &    \\
ring-id-not-dead &                 $-$         &    19                 &  5 &           - &  0 &                    - &  0 &     221 &  3 &                  &    \\
consensus-epr &                    $\checkmark$&    37                 &  5 &          57 &  5 &                   28 &  5 &       - &  0 &                  &    \\
client-server-ae &                 $\checkmark$&     4                 &  5 &          11 &  5 &                    4 &  5 &     442 &  5 &                  &    \\
client-server-db-ae &              $-$         &    16                 &  5 &          46 &  5 &                   37 &  5 &    6639 &  4 &                  &    \\
hybrid-reliable-broadcast &   $-$         &   178                 &  5 &           - &  0 &                    - &  0 &     937 &  5 &                  &    \\
paxos-epr &                        $\checkmark$&   920                 &  5 &       14332 &  4 &  568\footnotemark[1] &    &       - &  0 &                  &    \\
flexible-paxos-epr &               $\checkmark$&   418                 &  5 &        4928 &  5 &  561\footnotemark[1] &    &       - &  0 &                  &    \\
multi-paxos-epr &                  $\checkmark$&  4272                 &  4 &           - &  0 &                    - &  0 &       - &  0 &                  &    \\
stoppable-paxos-epr &              $\checkmark$&>18297                 &  2 &           - &  0 &                    - &  0 &       - &  0 &                  &    \\
fast-paxos-epr &                   $\checkmark$&  9630                 &  3 &           - &  0 &                    - &  0 &       - &  0 &                  &    \\
vertical-paxos-epr &               $\checkmark$&     -                 &  0 &           - &  0 &                    - &  0 &       - &  0 &                  &    \\
block-cache-async &                $-$         &     -                 &  0 &           - &  0 &                    - &  0 &       - &  0 &                  &    \\
bosco-3t-safety &                  $\checkmark$&>11019\footnotemark[2] &  1 &           - &  0 &                    - &  0 &       - &  0 &                  &    \\
            
\end{tabular}}
\end{center}\vspace{3pt}
\footnotemark[1]As reported in \cite{IC3PO-fmcad} on similar input.
\footnotemark[2]With EPR push down enabled.
\end{table*}

We present the results of our experiments in Table~\ref{tab:benchmark-results}. In general, for examples that converge with both prior approaches and our technique, we match or exceed existing results, with significant performance gains for some problems such as client-server-db-ae relative to the previous separation-based approach. Along with other techniques, we solve paxos-epr and flexible-paxos-epr, which are the simplest variants of Paxos in our benchmark, but nonetheless represents a significant jump in complexity over the examples solved by the prior generation of PDR/IC3 techniques. Paxos and its variants are notable for having invariants with two quantifier alternations ($\forall\exists\forall$) and a maximum quantifier depth of 6 or 7. We uniquely solve multi-, fast-, and stoppable-paxos-epr, which add significant complexity in the number of sorts, symbols, and quantifier depth required. Due to variations in seeds and the non-determinism of parallelism, our technique was only successful in some trials, but these results nevertheless demonstrate that our technique is capable of solving these examples. Our algorithm is unable to solve vertical-paxos-epr, as this example requires a 7 quantifier formula that is very expensive for our IG solver.

For universal-only examples, our algorithm is able to solve all but one of the examples\footnote{fast-paxos-forall, which is solved by our technique in the ablation study, albeit rarely.} solved by other techniques, and is able to solve one that others cannot. In some cases (e.g. consensus-forall), our solution is slower than other approaches, but on the whole our algorithm is competitive in a domain it is not specialized for. In addition, we significantly outperform the existing separation-based algorithm (fol-ic3) by solving several difficult examples (cache, paxos-forall).

\subsection{Ablation Study}
\begin{table*}[t]
\caption{Ablation study. Columns are interpreted as in \Cref{tab:benchmark-results}.}
\label{tab:ablation-study}
\centering
{\scriptsize
\setlength\tabcolsep{1pt}
\begin{tabular}{l|rrrrrrrrrrrrrrrrrr}
    Example &                  \bfseries{Our} & \# & No pDNF & \# & No EPR & \# & No Inc. SMT & \# & No Gen. & \# \\\hline
    lockserv &                             19 &  5 &         &    &        &    &     34 &  5 &      13 &  5 \\
    toy-consensus-forall &                  4 &  5 &         &    &        &    &      5 &  5 &       4 &  5 \\
    ring-id &                               7 &  5 &         &    &        &    &     11 &  5 &      13 &  5 \\
    sharded-kv &                            8 &  5 &         &    &        &    &     11 &  5 &       7 &  5 \\
    ticket &                               23 &  5 &         &    &        &    &     42 &  5 &      21 &  5 \\
    learning-switch &                      76 &  5 &         &    &        &    &    338 &  5 &     288 &  5 \\
    consensus-wo-decide &                  50 &  5 &         &    &        &    &     50 &  5 &      51 &  5 \\
    consensus-forall &                   1908 &  5 &         &    &        &    &   2154 &  5 &     558 &  5 \\
    cache &                              2492 &  4 &         &    &        &    & >16826 &  2 &   13116 &  5 \\
    paxos-forall &                        885 &  5 &         &    &        &    &   1071 &  5 &   10488 &  4 \\
    flexible-paxos-forall &              1961 &  5 &         &    &        &    &   1014 &  5 &   >4168 &  2 \\
    stoppable-paxos-forall &             7779 &  5 &         &    &        &    &   2820 &  5 &  >18727 &  1 \\
    fast-paxos-forall &                     - &  0 &         &    &        &    & >16573 &  1 &       - &  0 \\
    vertical-paxos-forall &                 - &  0 &         &    &        &    &      - &  0 &       - &  0 \\\hline
    firewall &                              4 &  5 &       4 &  5 &        &    &      4 &  5 &       4 &  5 \\
    sharded-kv-no-lost-keys &               4 &  5 &       4 &  5 &      4 &  5 &      5 &  5 &       4 &  5 \\
    toy-consensus-epr &                     4 &  5 &       5 &  5 &      5 &  5 &      5 &  5 &       5 &  5 \\
    ring-id-not-dead &                     19 &  5 &      37 &  5 &        &    &     44 &  5 &      52 &  5 \\
    consensus-epr &                        37 &  5 &     126 &  5 &    724 &  5 &     45 &  5 &     233 &  5 \\
    client-server-ae &                      4 &  5 &       3 &  5 &      4 &  5 &      4 &  5 &       4 &  5 \\
    client-server-db-ae &                  16 &  5 &      13 &  5 &        &    &     20 &  5 &      10 &  5 \\
    hybrid-reliable-broadcast &           178 &  5 &      98 &  5 &        &    &    173 &  5 &     629 &  5 \\
    paxos-epr &                           920 &  5 &   10135 &  4 &  >2895 &  1 &    609 &  5 &    3201 &  5 \\
    flexible-paxos-epr &                  418 &  5 &   13742 &  3 &      - &  0 &    775 &  5 &     799 &  5 \\
    multi-paxos-epr &                    4272 &  4 &  >15176 &  1 &      - &  0 &  15854 &  3 &    7326 &  4 \\
    stoppable-paxos-epr &              >18297 &  2 &       - &  0 &      - &  0 & >20659 &  1 &  >11946 &  1 \\
    fast-paxos-epr &                     9630 &  3 &       - &  0 &      - &  0 &   8976 &  3 &  >20871 &  2 \\
    vertical-paxos-epr &                    - &  0 &       - &  0 &      - &  0 &      - &  0 &       - &  0 \\
    block-cache-async &                     - &  0 &       - &  0 &        &    &      - &  0 &  >20038 &  2 \\
    bosco-3t-safety &                  >11019 &  1 &       - &  0 &      - &  0 &  >8581 &  1 &  >16689 &  1 \\
    
\end{tabular}}
\end{table*}

\begin{table*}[t]
\caption{Parallel vs sequential comparison. Each of 5 trials ran with 3 or 48 hour timeouts, respectively. The number of successes, and the average number of IG queries in each trial (including failed ones) are given. %Multi-block generalization, heuristic pushing, and prefix restrictions are disabled to allow comparing IG queries directly.
% Correct Prefix gives the mean total time spent in IG queries of the prefix.
}
\label{tab:par-seq-comparison}
\centering
{\scriptsize
\setlength\tabcolsep{4pt}
\begin{tabular}{l|rr|rr|rr}
                   & \multicolumn{2}{c|}{Successes} & \multicolumn{2}{c}{IG Queries} \\
Example &            Par. & Seq. & Par. & Seq.\\\hline
paxos-epr &             5 &    5 &   61 &   76\\
flexible-paxos-epr &    5 &    5 &   64 &   72\\
multi-paxos-epr &       3 &    1 &   67 &   84\\

\end{tabular}}
\end{table*}
\Cref{tab:ablation-study} presents an ablation study investigating effect of various features of our technique. The first column of \Cref{tab:ablation-study} repeats the full algorithm results, and the remaining columns report the performance with various features disabled individually.
The most important individual contributions come from $k$-pDNF matrices and EPR.
Using a 5-clause CNF instead of pDNF matrix (No pDNF) causes many difficult examples to fail and some (e.g., flexible-paxos-epr) to take significantly longer even when they do succeed.\footnote{With a single clause, there is no difference between CNF and $k$-pDNF so results are only given for existential problems.} Similarly, using full FOL mode instead of EPR (No EPR) leads to timeouts for all but the simplest Paxos variants.
Incremental SMT queries (No Inc. SMT) make the more difficult Paxos variants, and the universal cache example, succeed much more reliably. Multi-block generalization (No Gen.) makes many problems faster or more reliable, but disabling it allows block-cache-async to succeed.

To isolate the benefits of parallelism, we ran several examples in both parallel and serial mode with a proportionally larger timeout (\Cref{tab:par-seq-comparison}). In both modes we use a single prefix category containing all prefixes, with the same static order over prefixes.\footnote{To make the comparison cleaner, we also disabled multi-block generalization.}
Beyond the wall-clock speedup, the parallel IG algorithm affects the quality of the learned lemmas, that is, how well they generalize and avoid overfitting.
To estimate the quality of generalization, we count the total number of IG queries performed by each trial and report the average over the five trials. 
In all examples, the parallel algorithm learns fewer lemmas overall, which suggests it generalizes better. We attribute this improved generalization to the implicit bias towards lemmas that are faster to discover. For the more complicated example (multi-paxos-epr), this difference has an impact on the success rate.

\section{Related Work}\label{sec:related-work}

\paragraph{Extensions of PDR/IC3.}
The PDR/IC3~\cite{IC3,PDR} algorithm has been very influential as an invariant inference technique, first for hardware (finite state) systems and later for software (infinite state).
There are multiple extensions of PDR/IC3 to infinite state systems using SMT theories~\cite{DBLP:conf/sat/HoderB12,DBLP:journals/fmsd/KomuravelliGC16}.
\cite{UPDR} extended PDR/IC3 to universally quantified first-order formulas using the model-theoretic notion of \emph{diagrams}. \cite{DBLP:conf/atva/GurfinkelSV18} applies PDR/IC3 to find universally quantified invariants over arrays and also to manage quantifier instantiation.
Another extension of PDR/IC3 for universally quantified invariants is~\cite{DBLP:conf/sosp/MaGJKKS19},
where a quantified invariant is generalized from an invariant of a bounded, finite system. This technique of generalization from a bounded system has also been extended to quantifiers with alternations~\cite{IC3PO-NFM}. Recently, \cite{DBLP:conf/vmcai/ZhangGM21} suggested combining synthesis and PDR/IC3, but they focus on word-level hardware model checking and do not support quantifier alternations.
Most of these works focus on quantifier-free or universally quantified invariants. In contrast, we address unique challenges that arise when supporting lemmas with quantifier alternations.

The original PDR/IC3 algorithm has also been extended with techniques that use different heuristic strategies to find more invariants by considering additional proof goals and collecting reachable states~\cite{BetterGen,PttT}. Our implementation benefits from some of these heuristics, but our contribution is largely orthogonal as our focus is on inductive generalization of quantified formulas.
Generating lemmas from multiple states, similar to multi-block generalization, was explored in~\cite{DBLP:conf/cav/KrishnanCSG20}.

\cite{DBLP:conf/fmcad/MarescottiGHS17} suggests a way to parallelize PDR/IC3 by combining a portfolio approach with
problem partitioning and lemma sharing. Our parallelism is more tightly coupled into PDR/IC3,
as we parallelize the inductive generalization procedure.

\paragraph{Quantified Separation.}
Quantified separation~\cite{PLDI20} was recently introduced as a way to find quantified invariants with quantifier alternations. While~\cite{PLDI20} introduced a way to combine separation and PDR/IC3, it has limited scalability and cannot find the invariants of complex protocols such as Paxos.
Our work here is motivated by these scalability issues.
In contrast to~\cite{PLDI20}, our technique is able to find complex invariants by avoiding expensive but useless areas of the search space using a breadth-first strategy and a multi-dimensional inductive bias.
While \cite{PLDI20} searches for quantified lemmas in CNF, we introduce and use $k$-term pDNF. $k$-term pDNF can express the necessary lemmas of many distributed protocols more succinctly, resulting in better scalability.

\paragraph{Synthesis-Based Approaches to Invariant Inference.}
Synthesis is a common approach for automating invariant inference.
ICE~\cite{CAV14ice} is a framework for learning inductive invariants from positive, negative, and implication constraints. Our use of separation is similar, but it is integrated into PDR/IC3's inductive generalization, so unlike ICE we find invariants incrementally. 

\paragraph{Enumeration-Based Approaches.}
Another approach is to use enumerative search, for example
\cite{DBLP:conf/cav/FedyukovichPMG19}, which only supports universal quantification. Enumerative search has been extended to quantifier alternations in \cite{NSDI21}, which is able to infer the invariants of complex protocols such as some Paxos variants.

\section{Conclusion}\label{sec:conclusion}
We have presented an algorithm for quantified invariant inference that combines separation and inductive generalization. Our algorithm uses a breadth-first strategy to avoid regions of the search space that are expensive.
We also explore a new syntactic form that is well-suited for lemmas with alternations.
We show via a large scale experiment that our algorithm advances the state of the art in quantified invariant inference with alternations, and finds significantly more invariants on difficult problems than prior approaches.

\bibliographystyle{splncs04}
\bibliography{citations}

\end{document}